# Strong self-induced nonreciprocal transmission by using nonlinear PT-symmetric epsilon-near-zero metamaterials

Boyuan Jin[a], Christos Argyropoulos*[a]

[a]Department of Electrical and Computer Engineering, University of Nebraska-Lincoln, Lincoln, NE, USA 68588

*christos.argyropoulos@unl.edu

## ABSTRACT

Nonreciprocal transmission is the fundamental process behind unidirectional wave propagation phenomena. In our work, a compact and practical parity-time (PT) symmetric metamaterial is designed based on two Silicon Carbide (SiC) media separated by an air gap and photonically doped with gain and loss defects. We demonstrate that an exceptional point (EP) is formed in this PT-symmetric system when SiC operates as a practical epsilon-near-zero (ENZ) material and by taking into account its moderate optical loss. Furthermore and even more importantly, strong self-induced nonreciprocal transmission is excited due to the nonlinear Kerr effect at a frequency slightly shifted off the EP but without breaking the PT-symmetric phase. The transmittance from one direction is exactly unity while the transmittance from the other direction is decreased to very low values, achieving very high optical isolation. The proposed active nonlinear metamaterial overcomes the fundamental physical bounds on nonreciprocity compared with a passive nonlinear nonreciprocal resonator. The strong self-induced nonreciprocal transmission arises from the extreme asymmetric field distribution achieved upon excitation from opposite incident directions. The significant enhancement of the electric field in the defects effectively decreases the required optical power to trigger the presented nonlinear response. This work can have a plethora of applications, such as nonreciprocal ultrathin coatings for the protection of sources or other sensitive equipment from external pulsed signals, circulators, and isolators.

## 1. INTRODUCTION

Nonreciprocal transmission forms the basic operation mechanism of optical diodes, switches, and isolators and requires the breaking of Lorentz reciprocity law [1-3]. There are several ways to break reciprocity, such as exciting nonlinearity in passive asymmetric resonators [4, 5], applying magnetic material bias [6, 7] or dynamic space and time modulation [8, 9]. However, the passive nonlinear resonators are limited by fundamental physical bounds on nonreciprocity [10, 11]. In addition, magnetic materials are usually bulky, lossy, costly, and difficult to be integrated on chip. Furthermore, space and time modulation requires electro-optic converters which are challenging to be designed in high frequencies [12].

Another emerging field in optics and photonics is Parity-time (PT)-symmetry which corresponds to systems where gain and loss parameters are balanced in space, i.e., the refractive indices satisfy the relation $n(\vec{r}) = n^*(-\vec{r})$ for any position $\vec{r}$ in space [13]. The exceptional point (EP) is a special frequency point in PT-symmetric systems, where the effective non-Hermitian Hamiltonian becomes defective and the system exhibits unidirectional reflectionless transparency, where the transmittance is reciprocal and equal to unity and the reflectance only from one direction is equal to zero [14]. With the increase of the input laser power, the Kerr effect would inevitably alter the effective permittivity of the used materials and subsequently the optical response of the system. Since PT-symmetric systems are asymmetric, usually composed of a sequence of gain and loss components, the transmission can become nonreciprocal by triggering the optical nonlinearity close to the EP. Meanwhile PT-symmetric structures are active systems made of gain materials and can potentially overcome the fundamental physical bounds on nonreciprocity of any passive nonlinear nonreciprocal resonator system. Furthermore, nonlinear optical systems have ultrafast response times.

In this work, a compact and practical PT-symmetric metamaterial is proposed based on two Silicon Carbide (SiC) media separated by an air gap and photonically doped with gain and loss defects. The presented self-induced nonreciprocal structure is suitable for free-space optics applications. SiC has an epsilon-near-zero (ENZ) material response at mid-infrared (IR) frequencies with low optical loss. The used lossy ENZ medium is detrimental to the PT-symmetry as it breaks the balance between the gain and loss parameters in the system. Hence, the refractive indices of the system do not satisfy the relation $n(\vec{r}) = n^*(-\vec{r})$ anymore. However, we demonstrate that an EP is also formed in the proposed system by decreasing the thickness of the lossy ENZ medium, which mitigates the induced power loss. Even more importantly, we demonstrate strong self-induced nonreciprocal transmission at mid-IR excited by the Kerr nonlinear effect at a frequency slightly shifted off the EP but without breaking the PT-symmetric phase. The transmittance from one direction is exactly one while the transmittance from the other direction is decreased to very low values, achieving ultrahigh optical isolation. The strong self-induced nonreciprocal transmission arises from the extreme asymmetric field distribution achieved upon excitation from opposite incident directions. The significant enhancement of the electric field values in the defects effectively decreases the required optical power to trigger the presented nonlinear response. The low power can reduce the ohmic heating, and thus prevent the proposed nonreciprocal metamaterial from potential thermal damage. This work can pave the way to the design of a plethora of new compact optical components, such as nonreciprocal ultrathin coatings for the protection of sources or other sensitive equipment from external pulsed signals, circulators, and isolators.

## 2. EP FORMATION WITH LOSSY ENZ MEDIA IN THE LINEAR REGIME

The geometry of the proposed PT-symmetric metamaterial is shown in Fig. 1. Two identical ENZ media with thickness $b$ are separated by an ultrathin air gap $d$. Cylindrical defects with radius $r$ are located at the center of each ENZ segment and patterned along the y-axis with periodicity $a$. In the following, we normalize all the dimensions to $\lambda_{ENZ}$, i.e., the wavelength where the permittivity of the ENZ medium becomes equal to zero: $\text{Re}[\varepsilon_{ENZ}] \to 0$. The defects are doped with gain in one ENZ medium and with equal value loss in the other ENZ medium with permittivities satisfying the relation: $\varepsilon_{Loss} = \varepsilon_{Gain}^* = 4 - \delta i$. The ENZ media and defects are both non-magnetic with relative magnetic permeabilities $\mu_{ENZ} = \mu_d = 1$. The input waves are injected either from the forward or the backward directions, as shown in Fig. 1. The forward or backward inputs are both normal incident plane waves polarized along the y-axis with operation wavelength $\lambda = \lambda_{ENZ}$ and input intensity $I_0$

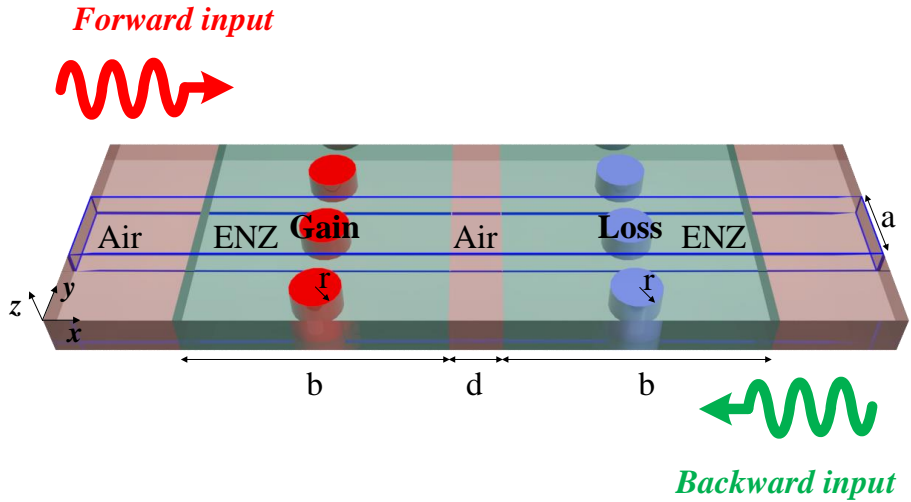

Figure 1. Schematic illustration of the proposed PT-symmetric metamaterial. Two ENZ media are separated by an air gap and loaded with cylindrical gain and loss defects, respectively. The blue frame shows the unit cell of the proposed metamaterial.

We start our investigation from the linear case where the input intensity $I_0$ is very low. When the ENZ medium is lossless, the effective permittivity and permeability of the composite ENZ metamaterial with defect is given by [15]:

$$\varepsilon_{eff} = \varepsilon_{ENZ} \to 0 \tag{1}$$

$$\mu_{eff} = \mu_0 \left( \frac{A_e}{A_t} + \frac{2\pi r J_1(k_d r)}{A_t k_d J_0(k_d r)} \right), \tag{2}$$

where $A_t = a \cdot b$ is the total area of the ENZ and defect, $\varepsilon_d$ and $k_d$ are the complex permittivity and wave vector in the defect, $A_e = A_t - A_d$ is the ENZ material area, $A_d = \pi r^2$ is the defect area, and $J_n(x)$ is the n-th order Bessel function. The EP occurs when $\text{Re}[\mu_{eff}] \to 0$ and $\text{Im}[\mu_{eff}] = \pm \lambda_{ENZ} / \pi b$ [15]. Therefore, by tuning the structure's dimensions, an EP is obtained with the proposed structure at $\lambda_{ENZ}$ when $a = \lambda_{ENZ}$, $b = 3.27\lambda_{ENZ}$, $r = 0.19\lambda_{ENZ}$, $d = 0.25\lambda_{ENZ}$, $\varepsilon_{ENZ} = 1 \times 10^{-4}$ and $\delta = 0.01$. Due to the resonant feature introduced by the air gap, we can achieve a series of EPs by using different values of $d$ equally spaced with a step size of $\lambda_{ENZ} / 2$ [16]. The EP is verified in Fig. 2, where the linear transmission and reflection spectra of the proposed metamaterial are computed by using COMSOL Multiphysics, a commercial full-wave simulation software. The dispersion of the materials is neglected, since we focus our study only in a narrow wavelength range around $\lambda_{ENZ}$. It can be seen that the forward and backward transmittances are reciprocal and equal to unity $(T_F = T_B = 1)$ at $\lambda_{ENZ}$, while the forward reflectance is $R_F = 16$ and the backward reflectance is $R_B = 0$. This unidirectional reflectionless transparency observed at $\lambda_{ENZ}$ in Fig. 2 is a typical property of EPs [14].

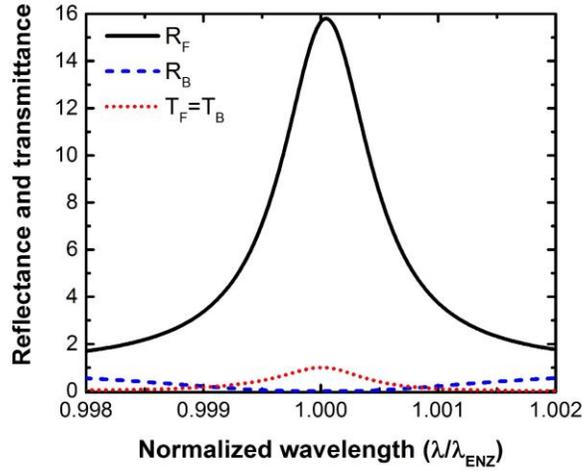

Figure 2. The computed spectra of the linear reflectance and transmittance of the proposed PT-symmetric metamaterial with $a = \lambda_{ENZ}$, $b = 3.27\lambda_{ENZ}$, $r = 0.19\lambda_{ENZ}$, $d = 0.25\lambda_{ENZ}$, and assuming lossless ENZ media. An EP occurs at $\lambda = \lambda_{ENZ}$ with $T_B = T_F = 1$, $R_B = 0$, and $R_F \neq 0$.

The ENZ response in real materials, such as transparent conductive oxides (TCOs) and SiC, is not ideal (lossless) due to a moderate optical loss [17-21]. Figure 3 shows the backward reflectance $R_B$ and transmittance $T_B$ when the ENZ media have a non-zero imaginary part of permittivity. The dimensions and other parameters are all the same to those used in Fig. 2. We can see that the optical loss of the ENZ media deteriorates the unidirectional reflectionless transparency and thus the obtained EP. Even a small imaginary value $(\text{Im}[\varepsilon_{ENZ}] = -0.001)$ at the ENZ permittivity can significantly decrease the unity transmittance and undermine the zero reflectance achieved at the EP. This is because the PT-symmetry relation $n(\vec{r}) = n^*(-\vec{r})$ is not strictly satisfied in the presented structure when the ENZ media are lossy.

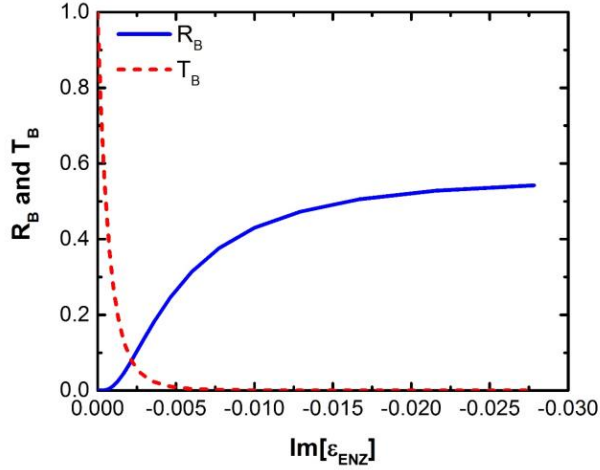

Figure 3. Backward reflectance and transmittance at $\lambda = \lambda_{ENZ}$ as functions of $\text{Im}[\varepsilon_{ENZ}]$. The optical loss of the ENZ media deteriorates the EP condition.

The power attenuation in the lossy ENZ material can be mitigated by reducing the thickness of the ENZ media. Consequently, the other dimensions and parameters must be changed to maintain the EP at $\lambda = \lambda_{ENZ}$. To provide further insights in this issue, we study again the PT-symmetric metamaterial made of lossless ENZ media. The dependence of $R_B$ and $T_B$ on the periodicity $a$ and the ENZ media thickness $b$ is shown in Fig. 4. The periodicity $a$ should be increased, while decreasing the ENZ media thickness $b$, in order to reach again the unity transmittance condition. On the other hand, the backward reflectance takes non-zero values at the unity transmittance points, as shown in Fig. 4(a). To further suppress the backward reflectance, the defects need to be doped with higher gain/loss parameters.

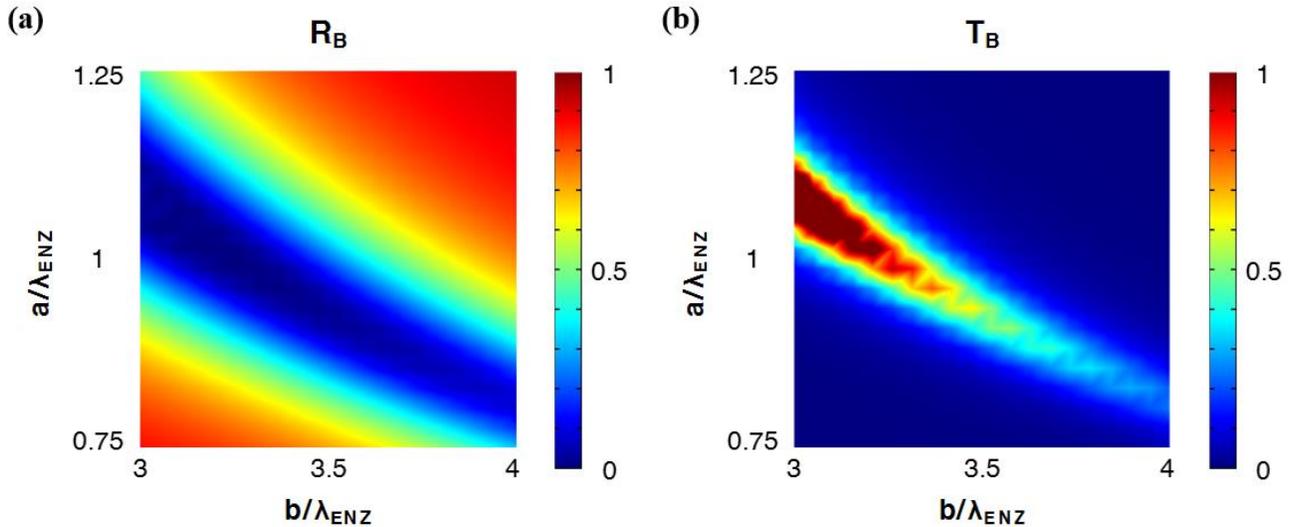

Figure 4. Backward (a) reflectance and (b) transmittance at $\lambda = \lambda_{ENZ}$ as functions of the metamaterial dimensions $a$ and $b$ in the case of lossless ENZ material.

In the results shown in Fig. 5, SiC is used as the ENZ material with $\text{Im}(\varepsilon_{ENZ}) = -0.03$ at its mid-IR ENZ response [19-21]. The thickness of the ENZ media is decreased to $b = \lambda_{ENZ}$, and the other dimensions are changed to $a = 2\lambda_{ENZ}$ and $d = 0.52\lambda_{ENZ}$. The computed linear reflectance and transmittance at the ENZ wavelength $\lambda = \lambda_{ENZ}$ as functions of $\delta = |\text{Im}[\varepsilon_d]|$ under both illuminating directions are shown in Fig. 5. The EP occurs at $\delta = 0.13$ with the unidirectional reflectionless transparency condition becoming the following: $T_F = T_B = 1$, $R_F = 6$, and $R_B = 0.037$. The required gain coefficient $\delta$ value is much larger compared to the lossless ENZ media but still in a practical feasible range [22, 23].

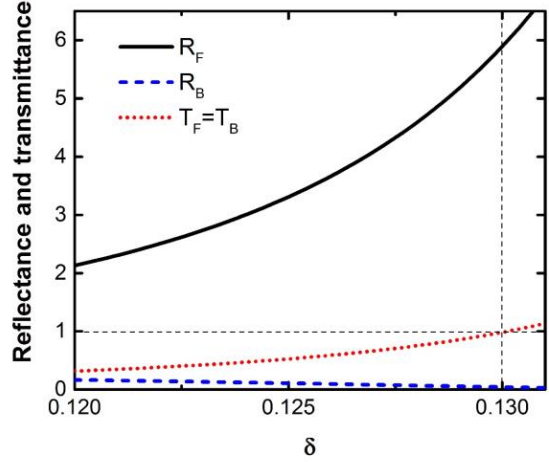

Figure 5. Linear reflectance and transmittance at $\lambda = \lambda_{ENZ}$ as functions of gain coefficient $\delta$, where $b = \lambda_{ENZ}$, $a = 2\lambda_{ENZ}$, $d = 0.52\lambda_{ENZ}$, and the ENZ material is SiC with $\text{Im}(\varepsilon_{ENZ}) = -0.03$.

## 3. NONRECIPROCAL TRANSMISSION BY KERR NONLINEARITY WITH LOSSY ENZ MEDIA

With the increase of the input power, the Kerr effect is inevitable, leading to the nonlinear permittivity $\varepsilon_{NL} = \chi^{(3)} |E|^2$, where $\chi^{(3)}$ is the third-order nonlinear susceptibility [22-26]. We can see that the Kerr effect can be boosted by using higher $\chi^{(3)}$ materials or larger optical input intensity values. For the cylindrical defects where the electric field is tightly confined and enhanced, we assume the doped materials to have a relative weak nonlinearity with $\chi_d^{(3)} = 6 \times 10^{-20}$ m²/V², a typical value of several dielectric materials [27]. In addition, we neglect the nonlinearity in the ENZ media, i.e., $\chi_{ENZ}^{(3)} = 0$, since the field is very weak at this region. The $\chi^{(3)}$ values used here can provide a conservative estimate of the input intensity required to obtain self-induced nonreciprocal transmission. Note that ENZ materials often have large nonlinear susceptibilities and the nonreciprocal transmission can occur at lower input intensities compared to the simulation results shown in the next paragraph [16].

After introducing the Kerr nonlinearity at $\lambda = \lambda_{ENZ}$, the forward $T_F$ and backward $T_B$ transmittances as functions of the periodicity $a$ and the input intensity $I_0$ are computed and plotted in Fig. 6. The parameters are the same to those used in Fig. 5 except for $d = 0.51\lambda_{ENZ}$, making the system to slightly deviate from the EP obtained in the linear regime. It can be seen that both $T_F$ and $T_B$ exhibit an abrupt jump from close to zero to a peak value when the parameters $a$ or $I_0$ are swept. The peak of $T_F$ occurs at low input intensities, while the peak of $T_B$ is obtained at relatively higher intensity values. The lagging between the peaks of $T_F$ and $T_B$ suggests that the proposed nonlinear PT-symmetric metamaterial can achieve nonreciprocal transmission at some particular $a$ and $I_0$ values.

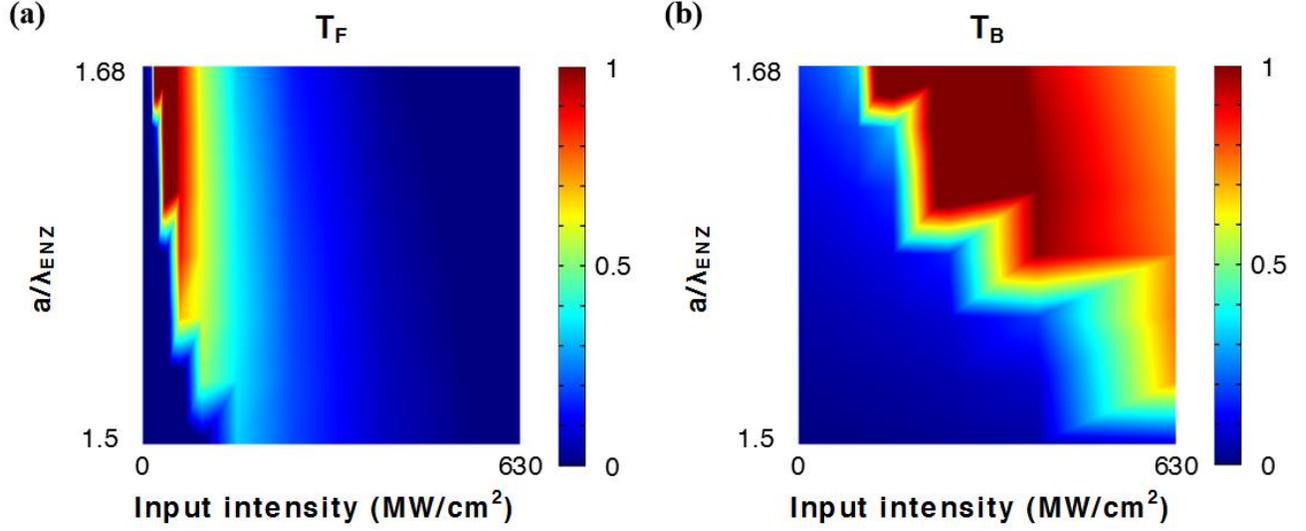

Figure 6. (a) Forward $T_F$ and (b) backward $T_B$ transmittances as functions of the input intensity $I_0$ and the periodicity $a$, where $d = 0.51\lambda_{ENZ}$ and $\lambda = \lambda_{ENZ}$

When $a = 1.6\lambda_{ENZ}$ and the other parameters are kept the same to those used in Fig. 6, the forward $T_F$ and backward $T_B$ transmittances as functions of the input intensity are shown in Fig. 7(a). At low input intensities, the system is linear, and the transmission is reciprocal and very close to zero under both illumination directions. The peaks of $T_F$ and $T_B$ are located at $I_0 = 50$ MW/cm$^2$ and $I_0 = 300$ MW/cm$^2$, respectively. The system exhibits significant self-induced nonreciprocal transmission at these two peaks. The maximum transmission contrast occurs at the $T_F$ peak with $T_F = 1$ and $T_B = 0.15$. The electric field distribution is computed and shown in Fig. 7(b), where a strong field enhancement in the active (gain) defect can be observed at $I_0 = 50$ MW/cm$^2$ only for the forward direction illumination.

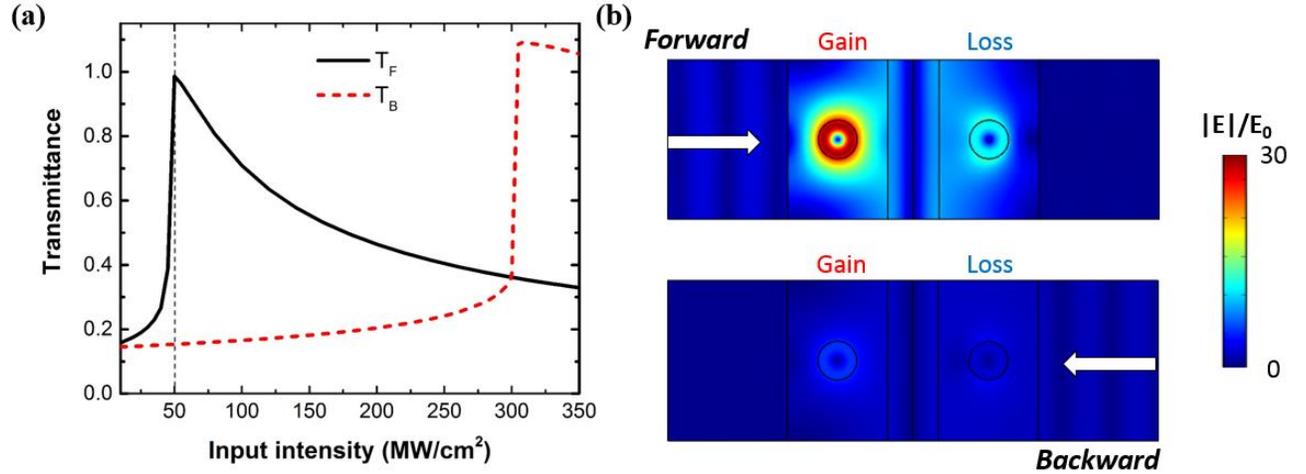

Figure 7. (a) Forward and backward transmittances as functions of the input intensity $I_0$, where $a = 1.6\lambda_{ENZ}$, $d = 0.51\lambda_{ENZ}$, and $\lambda = \lambda_{ENZ}$. (b) Computed distribution of the electric field enhancement at $\lambda_{ENZ}$ and $I_0 = 50$ MW/cm$^2$ under forward and backward illuminations, respectively. $E$ is the local electric field and $E_0$ is the electric field amplitude of the incident wave.

## 4. CONCLUSIONS

To conclude, a compact and practical parity-time (PT) symmetric metamaterial is proposed by using SiC as the ENZ material photonically doped with gain and loss defects. To mitigate the detrimental effects of the optical power loss in the lossy ENZ material, we decrease the thicknesses of the SiC segments. We demonstrate that the proposed structure made of lossy ENZ media support an EP. Furthermore and more interestingly, strong self-induced nonreciprocal transmission is observed by the proposed metamaterial when the nonlinear Kerr effect is triggered. The asymmetric and strong field enhancement in the gain/loss defects effectively decreases the required optical power to generate the presented nonlinear response. The dimensions of the proposed metamaterial can be further optimized to improve the isolation ratio. This work can have a plethora of new applications, such as nonreciprocal ultrathin coatings for the protection of sources or other sensitive equipment from external pulsed signals, circulators, and isolators.